\documentclass{Interspeech2024}

\usepackage{cite}
\usepackage{comment}
\usepackage{arydshln}

\interspeechcameraready

\title{Phonetic Error Analysis of Raw Waveform Acoustic Models\\with Parametric and Non-Parametric CNNs}

\name[affiliation={1,2}]{Erfan}{Loweimi}
\name[affiliation={3}]{Andrea}{Carmantini}
\name[affiliation={3}]{Peter}{Bell}
\name[affiliation={3}]{Steve}{Renals}
\name[affiliation={2}]{Zoran}{Cvetkovic}

\address{
  $^1$Speech Group, Machine Intelligence Laboratory, University of Cambridge, UK\\
  $^2$King's College London, UK\\
  $^3$Centre for Speech Technology Research (CSTR), University of Edinburgh, UK}
\email{}

\keywords{Raw waveform modelling, phone recognition, phonetic error analysis, broad phonetic class, confusion matrix}

\newcommand{\specialcell}[2][c]{%
\begin{tabular}[#1]{@{}c@{}}#2\end{tabular}}

\begin{document}

\maketitle

\begin{abstract}
In this paper, we analyse the error patterns of the raw waveform acoustic models in TIMIT's 
phone recognition task. Our analysis goes beyond the conventional phone error rate (PER) metric. We categorise the phones into three groups: \{affricate, diphthong, fricative, nasal, plosive, semi-vowel, vowel, silence\}, \{consonant, vowel$^+$, silence\}, and \{voiced, unvoiced, silence\} and, compute the PER for each broad phonetic class in each category.
We also construct a confusion matrix for each category using the substitution errors and compare the confusion patterns with those of the Filterbank and Wav2vec 2.0 systems.
Our raw waveform acoustic models consists of parametric (Sinc2Net) or non-parametric CNNs and Bidirectional LSTMs, achieving down to 13.7\%/15.2\% PERs on TIMIT Dev/Test sets, outperforming reported PERs for raw waveform models in the literature.
We also investigate the impact of transfer learning from WSJ on the phonetic error patterns and confusion matrices. It reduces the PER to 11.8\%/13.7\% on the Dev/Test sets.
\end{abstract}


\section{Introduction}
\label{sec:intro}

The conventional metric for evaluating phone recognition systems is phone error rate (PER), which measures the Levenshtein distance involving substitution, deletion, and insertion errors. However, PER lacks insight into the contribution of various broad phonetic classes (BPCs). 
In \cite{lea2023}, a detailed phonetic error analysis was carried out, computing the percentage of PER associated with each BPC. To this end, three phonetic categories were defined: \{affricate, diphthong, fricative, nasal, plosive, semi-vowel, vowel, silence\}, \{consonant, vowel$^+$, silence\}, and \{voiced, unvoiced, silence\}. Then, the substitution, deletion, insertion and subsequently the PER for each BPC within these categorisations were computed. A confusion matrix for each category was also calculated and the confusion patterns were analysed and visualised for various types of models.

In this paper, we perform phonetic error analysis using BPCs on the raw waveform acoustic models, in contrast to \cite{lea2023} where the acoustic models are Mel Filterbank-based. Raw waveform models perform minimal processing, leaving the speech parametrisation to be learned jointly with the acoustic model, tailored for the given task. As such there is no task-blind and lossy feature engineering process which may inadvertently lead to task-relevant information loss. Further, compared with the MFCC or Filterbank features, raw waveform models have access to information encoded in the Fourier transform's phase spectrum, demonstrated to be useful in a wide range of applications \cite{Myphdthesis}, including speech reconstruction \cite{INTERSPEECH2011}, recognition \cite{INTERSPEECH2023,ICASSP2021,ICASSP2017}, enhancement \cite{enhace-phase-2019,ICEE2011}, source-filter separation \cite{INTERSPEECH2017_ph}, etc. 

The key contributions of this paper are summarised below:

\begin{itemize}
    \item Development of raw waveform acoustic models with a cascade of parametric (Sinc2Net \cite{INTERSPEECH2019}) or non-parametric CNNs and recurrent layers, which achieve the highest performance on TIMIT \cite{TIMIT}, compared to other raw waveform models. 

    \item Calculation of the PER for all broad phonetic classes within each phonetic categorisation for the raw waveform models. 

    \item Computation of a confusion matrix for each phonetic categorisation for the raw waveform models.

    \item Exploration of the impact of transfer learning from WSJ \cite{WSJ} on the phonetic errors and confusion matrices.  

    \item Comparative analysis of the PER per BPC and the confusion patterns of the raw waveform models with the state-of-the-art Wav2vec 2.0 \cite{wav2vec2020} and Filterbank based systems.
\end{itemize}


Having reviewed the related work in Section~2, covering the raw waveform acoustic modeling and applications of the BPCs in speech processing, Section~3 describes the architecture of our raw waveform acoustic models. Section~4 details the three phonetic categorisations. Section~5 presents the experimental results as well as discussion and Section~6 concludes the paper.

\section{Related Work}
\label{sec:review}

\subsection{Raw waveform Acoustic Modelling}
Palaz et al \cite{PALAZ201915} investigated the usefulness of raw waveform models on the TIMIT phone recognition task and showed CNNs have superior performance over fully-connected networks. Tuske et al \cite{Tuske2018} compared raw waveform with traditional features in an LVCSR task. Sainath et al \cite{sainath2016ICASSP} deployed raw waveform modelling for joint acoustic modelling and beamforming in a multi-channel scenarios.
Ghahremani et al \cite{Ghahremani2016} used a TDNN architecture for raw waveform modeling and investigated the usefulness of i-vector for speaker adaptation.
~Zhu et al \cite{ZhuIS2016} and Von Platen et al \cite{Platen2019} built multi-scale raw waveform models, to construct representations with high spectral and temporal resolutions. 
Advantages of modelling speech in the waveform domain have been shown earlier also in the context of SVM and GMM-HMM approaches \cite{Yousafzai11a,Ager2011}.

The above cited works rely on conventional CNNs, which employ non-parametric FIR filters, while another line of research employs parameterised CNNs characterised by few parameters. SincNet \cite{sincnet1}, the first of this type of CNNs, have been applied for phone recognition \cite{sincnet1,e2e-sincnet2020}, speech recognition (both hybrid \cite{INTERSPEECH2020-wave} and end-to-end (E2E) \cite{e2e-sincnet2020}) and speaker recognition \cite{sincnet1}. The kernel of the SincNet's filters in the time domain, is a sinc function, leading to a filterbank with rectangular filters in the frequency domain. Each filter is characterised by two trainable parameters: centre frequency and bandwidth. 
Loweimi et al \cite{INTERSPEECH2019} generalised this idea to modulated kernel-based CNNs and developed Sinc2Net, GammaNet and GaussNet where the filters in the frequency domain take triangular, Gammatone and Gaussian shapes, respectively.
Other examples of parametric CNNs include ParzNet \cite{Oglic2021} and Complex Gabor CNN (CGCNN) \cite{cgcnn2020}. 
Yue et al \cite{INTERSPEECH2022} applied parametric CNNs in Dysarthric speech recognition. 
Fainberg et al \cite{FeinburgASRU2019} studied the 
speaker adaptation 
via retraining the Sinc layer parameters and showed this functionally resembles the VTLN. 


\subsection{Applications of BPCs}
The notion of broad phonetic classes, used in this work for phonetic error analysis, has had a wide range of applications. 
In speaker verification and identification tasks, BPCs --particularly vowels and nasals-- proved more informative than other broad phonetic classes \cite{Eatock1994,Margit2006}. Lu et al~\cite{Lu2020} showed that using BPCs' posteriorgrams can improve speech quality and intelligibility in the speech enhancement task. BPCs have also been applied in language identification \cite{Kempton2008} and in speech coding \cite{Zhang1997} by allocating different number of bits to speech frames belonging to different BPCs. 
They were also proven useful for speech emotion recognition \cite{Yuan2021TheRO}, particularly the vowel class. In addition, the BPCs were applied as a loss function in phone recognition \cite{Lee2019}, and in automatic speech recognition (ASR) for decision tree-based state clustering \cite{Young2006}, guiding the decoding process \cite{Ziegler2012} and multilingual speech recognition \cite{ZGANK2005379}.

\section{Architecture}
\label{sec:arch}
\begin{figure}[t]
  \centering
  \includegraphics[width=\linewidth]{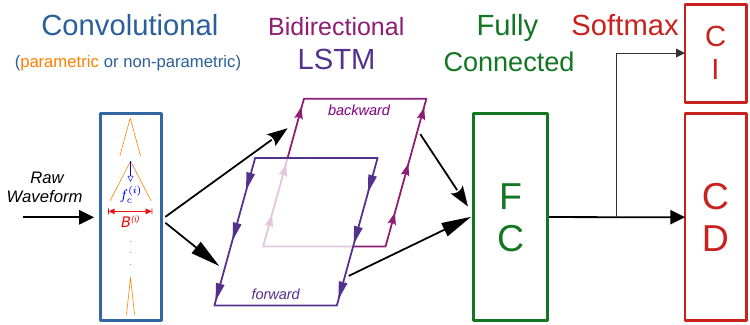}
  \caption{Our raw waveform acoustic models consist of a cascade of (parametric or non-parametric) convolutional, BLSTM and fully-connected (FC) layers. The output layer composed of context-dependent (CD) and context-independent (CI) heads.}
  \label{fig:arch}
\end{figure}

Fig.~\ref{fig:arch} depicts the architecture we employed for raw waveform acoustic modeling, consisting of a cascade of parametric or non-parametric convolutional, Bidirectional Long Short-Term Memory (BLSTM) \cite{Graves2005}, and fully-connected (FC) layers. This design leverages complementary modeling capabilities of individual layers: CNN for feature extraction, BLSTM for context and sequential modelling and FC layer(s) for further abstraction extraction and improvement of linear separability of the classes, right before the softmax layer which essentially is a linear classifier. The output layer comprises two heads: one for context-dependent (CD) state-clustered triphones and another for context-independent (CI) monophones. The CD head plays the key role and the CI is utilised for regularisation purposes.

We experimented with both parametric and non-parametric convolutional layers. For the former, we adopted Sinc2Net \cite{INTERSPEECH2019} whose 
kernel in the time domain is the Sinc-squared function

\begin{equation}
h^{(i)}(t) \, = \, \underbrace{\text{sinc}^2(B^{(i)} t)}_{\text{kernel}} \, \underbrace{\cos(2 \pi f^{(i)}_c t)}_{\text{carrier}}
\end{equation}

\noindent where $h^{(i)}(t)$, $B^{(i)}$ and $f^{(i)}_c$ denote the impulse response, bandwidth and centre frequency of the $i^{th}$ filter, respectively.

In the frequency domain, Sinc2Net acts as a filterbank with triangular filters centred around corresponding carrier frequencies, and is thus closely comparable with the triangular filters used in the Mel Filterbank
~(FBank) features. The triangular filters 
~are biologically more plausible than the rectangular filters in SincNet as they implicitly model the spectral masking \cite{moore2004psychology}.

Here, both CNN and FC sub-networks contain one layer. The FC layer includes 1024 nodes and the convolutional one consists of 128 kernels of length 129 with a max pooling of size 4. Dropout \cite{Srivastava2014} and ReLU activation function are used in both convolutional and FC layers. The BLSTM layers contain 550 nodes in each direction along with dropout. Batch normalisation \cite{batchnorm2015} was used in both BLSTM and FC layers. 

\section{Phonetic Categorisations}
\label{sed:bpc}

We have used three phonetic categorisations, similar to \cite{lea2023}, specified in Table~\ref{tab:bpc}.
Note that silence in all categorisations remains identical and encompasses non-speech segments at the beginning/end of utterances, epenthetic silence \cite{TIMIT}, short pauses, and closures before the Plosives. Additionally, the Vowel$^+$ in the second category, represents the union of vowels and diphthongs, grouped together due to their similarity \cite{lea2023}.

The sum of the PERs of all broad phonetic classes ($c$) within each category ($C$) equals the overall PER:
\begin{equation}
    \text{PER} \ = \ \sum_{c \in C} \text{PER}_c \ \stackrel{\text{e.g.}}{= } \ \text{PER}_\text{voi} + \text{PER}_\text{sil} + \text{PER}_\text{unv} 
\end{equation}
\noindent For example, the overall PER equals the sum of PERs of the Voiced, Silence and Unvoiced BPCs.

\section{Experimental Results and Discussion}
\label{sec:exp}

Models were trained using the PyTorch-Kaldi toolkit \cite{pytorch-kaldi,Kaldi2011} with the cross entropy loss and batch size of 8. The CD and CI output heads consist of 1936 and 48 nodes, respectively. The FBank features are 83-D: 80 filters plus three pitch-related features. 
For the transfer learning from WSJ, systems were initially trained on WSJ, and then only the weights between the penultimate and output layers were trained from scratch on TIMIT.

\begin{table}[t!]
\centering
\caption{{Mapping of phones to BPCs in three categorisations.}}
\vspace{-2mm}
\begin{tabular}{l|c}
\hline
         classes & phones  \\
\hline
\hline
\\[-3mm]
Affricates (aff)  & ch jh  \\
Diphthongs (dip)  & aw ay ey ow oy \\
Fricatives (fri)  & dh f s sh th v z \\
Nasal (nas)       & m n ng \\
Plosive (plo)     & b d dx g k p t \\
Semi-vowel (sem)  & hh l r w y \\
Silence (sil)     & sil \\
Vowel (vow)       & aa ae ah eh er ih iy uh uw \\
\hline
Consonant (con)     & \specialcell{b ch d dh dx f g hh jh k l m n ng p r s sh\\t th v w y z}  \\
Silence (sil)       & sil \\
Vowel$^+$ (vow$^+$) & aw ay ey ow oy aa ae ah eh er ih iy uh uw \\
\hline
Voiced (voi)    & \specialcell{aa ae ah aw ay b d dh dx eh er ey g hh\\ih iy jh l m n ng ow oy r uh uw v w y z} \\ 
Silence (sil)   & sil \\
Unvoiced (unv)  & ch f k p s sh t th \\
\hline
\hline
\end{tabular}
\label{tab:bpc}
\end{table}

\begin{table}[t!]
\centering
\caption{{PERs of various phone recognition systems on TIMIT.}}
\vspace{-2mm}
\begin{tabular}{l|c|cc}
\hline
 Feature  & Architecture & Dev & Test  \\
\hline
\hline
FBank-83 \cite{lea2023} & Best System in \cite{lea2023} & 12.8 & 14.1  \\
FBank-83-WSJ \cite{lea2023} & Best System in \cite{lea2023} & 11.5 & 13.1 \\
\hline
Raw-Wav \cite{PALAZ201915} & CNN & - & 21.9 \\
Raw-Wav (E2E) \cite{e2e-sincnet2020}  & CNN & 18.9 & 21.1  \\
Raw-Wav (E2E) \cite{e2e-sincnet2020}  & SincNet & 17.3 & 19.3  \\
Raw-Wav \cite{sincnet1}  & CNN & - & 18.1 \\
Raw-Wav \cite{sincnet1}  & SincNet & - & 17.2  \\
Raw-Wav \cite{INTERSPEECH2019} & GammaNet & - & 17.2 \\
Raw-Wav \cite{cgcnn2020}  & CGCNN & 15.2 & 17.1  \\
Raw-Wav \cite{INTERSPEECH2019} & GaussNet & - & 17.0 \\
Raw-Wav \cite{INTERSPEECH2019} & Sinc2Net & - & 16.9 \\
Raw-Wav \cite{Oglic2021} & ParzNet & 15.0 & 16.5 \\
Raw-Wav \cite{Loweimi2023-real} & CNN & 14.9 & 16.5 \\
\hline
Raw-Wav & Sinc2Net & 13.7 & 15.5 \\   
Raw-Wav & CNN & 14.1 & 15.2  \\   
Raw-Wav-WSJ & CNN & 11.8 & 13.7 \\
\hline
\hline
\end{tabular}
\label{tab:per}
\end{table}

Table~\ref{tab:per} presents the PER on TIMIT's Dev and Test sets, comparing the performance of various raw waveform systems reported in the literature. As seen, the performance of the proposed raw waveform models, with both parametric and non-parametric CNNs, are close to each other and outperform other models with a notable margin. This performance gain is primarily due to the effective combination of the CNNs and BLSTMs.

\begin{figure}[t!]
  \centering
  \includegraphics[width=\linewidth]{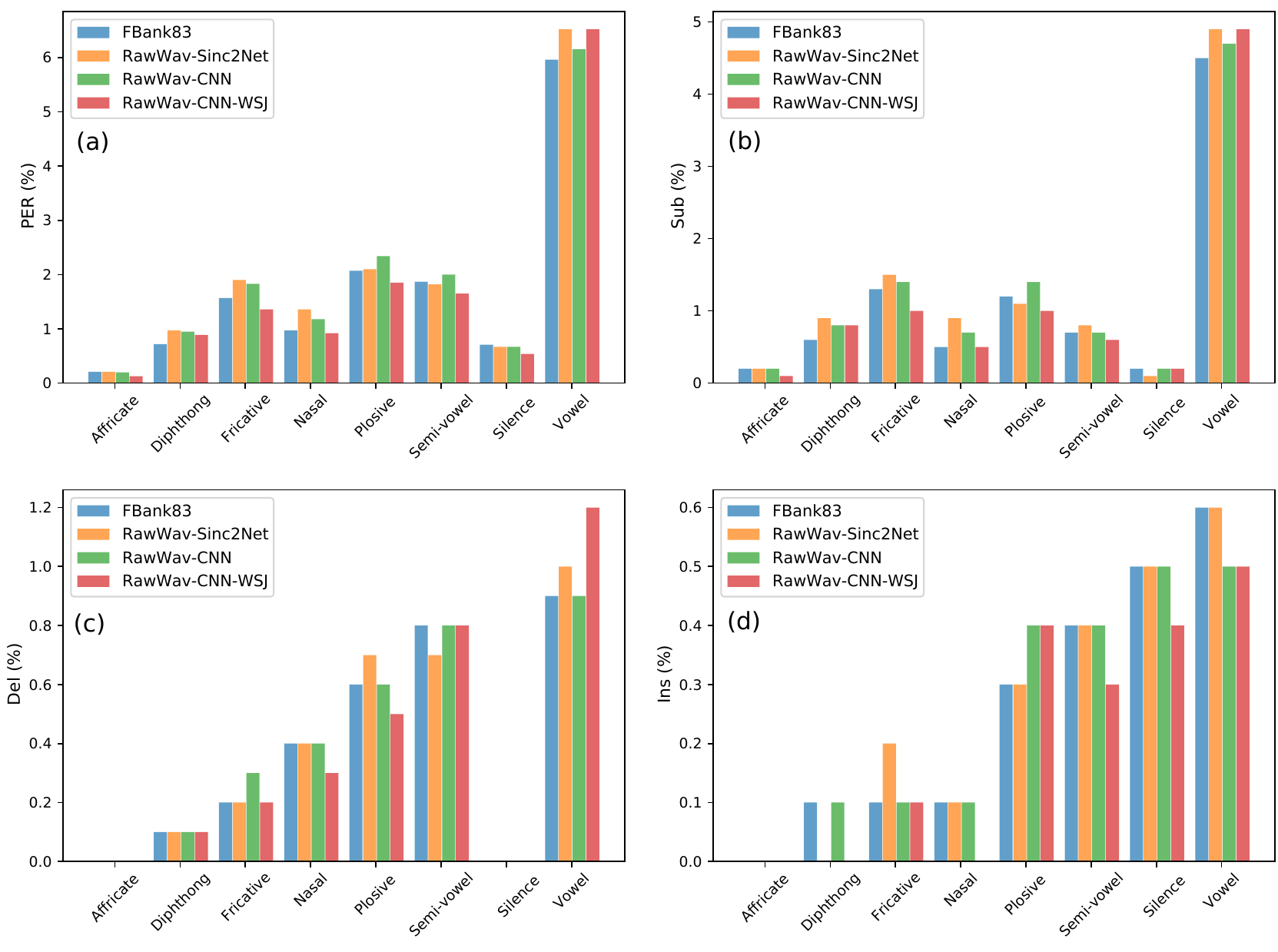}
  \vspace{-5mm}
  \caption{Recognition errors for FBank and raw waveform models. (a) PER, (b) Sub, (C) Del, (e) Ins.}
  \label{fig:psdi}
\end{figure}

\begin{figure}[t!]
  \centering
  \includegraphics[width=\linewidth]{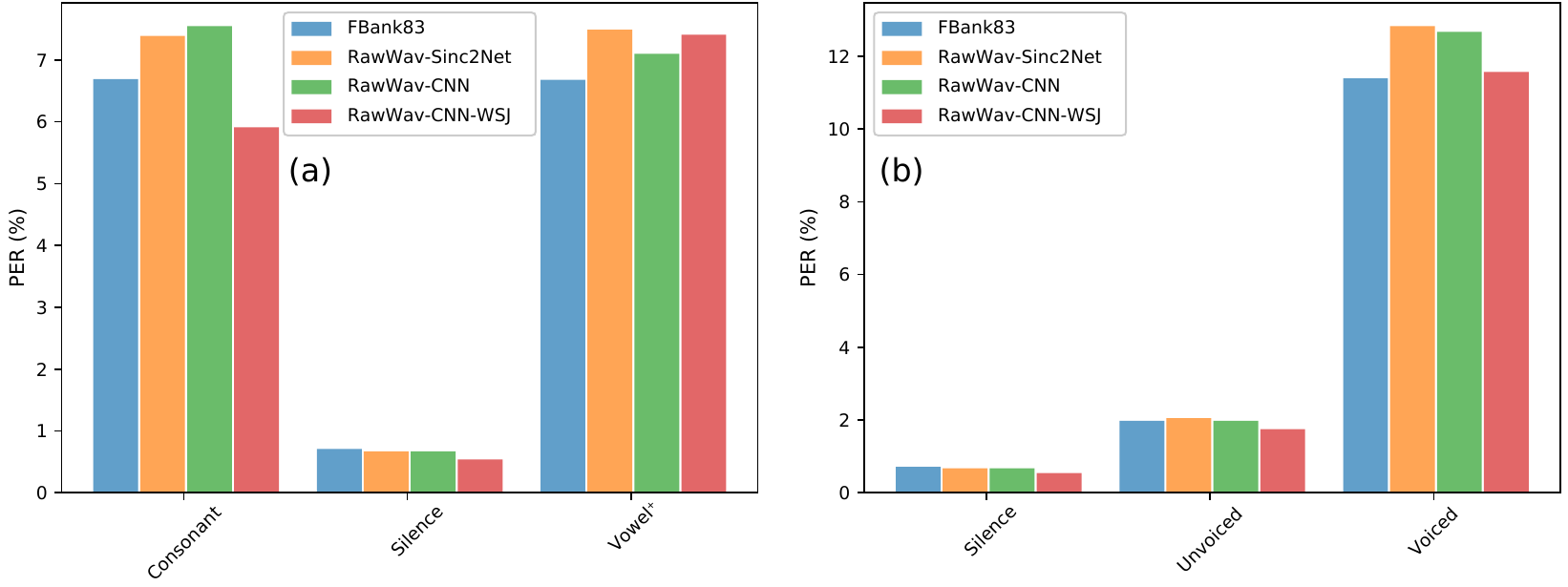}
  \vspace{-4mm}
  \caption{PER for FBank and raw waveform models. (a) Consonant/Vowel$^+$/Silence, (b) Voiced/Unvoiced/Silence.}
  \label{fig:cv-vu}
\end{figure}

Figs.~\ref{fig:psdi} and \ref{fig:cv-vu} depict the breakdown of PER over the Filterbank and raw waveform models. Despite differences, the overall trends in phonetic error distribution remain consistent across these systems. For example, the largest errors always belong to the vowel class, due to being highly sensitive to the speaker attributes such as ID \cite{Eatock1994,Margit2006} and emotion \cite{Yuan2021TheRO}. 
The consistent trends observed across various front-end and back-end configurations imply that the fundamental challenge of class confusions transcends the specific choices of these components. 

\begin{figure}[t!]
  \centering
  \includegraphics[width=\linewidth]{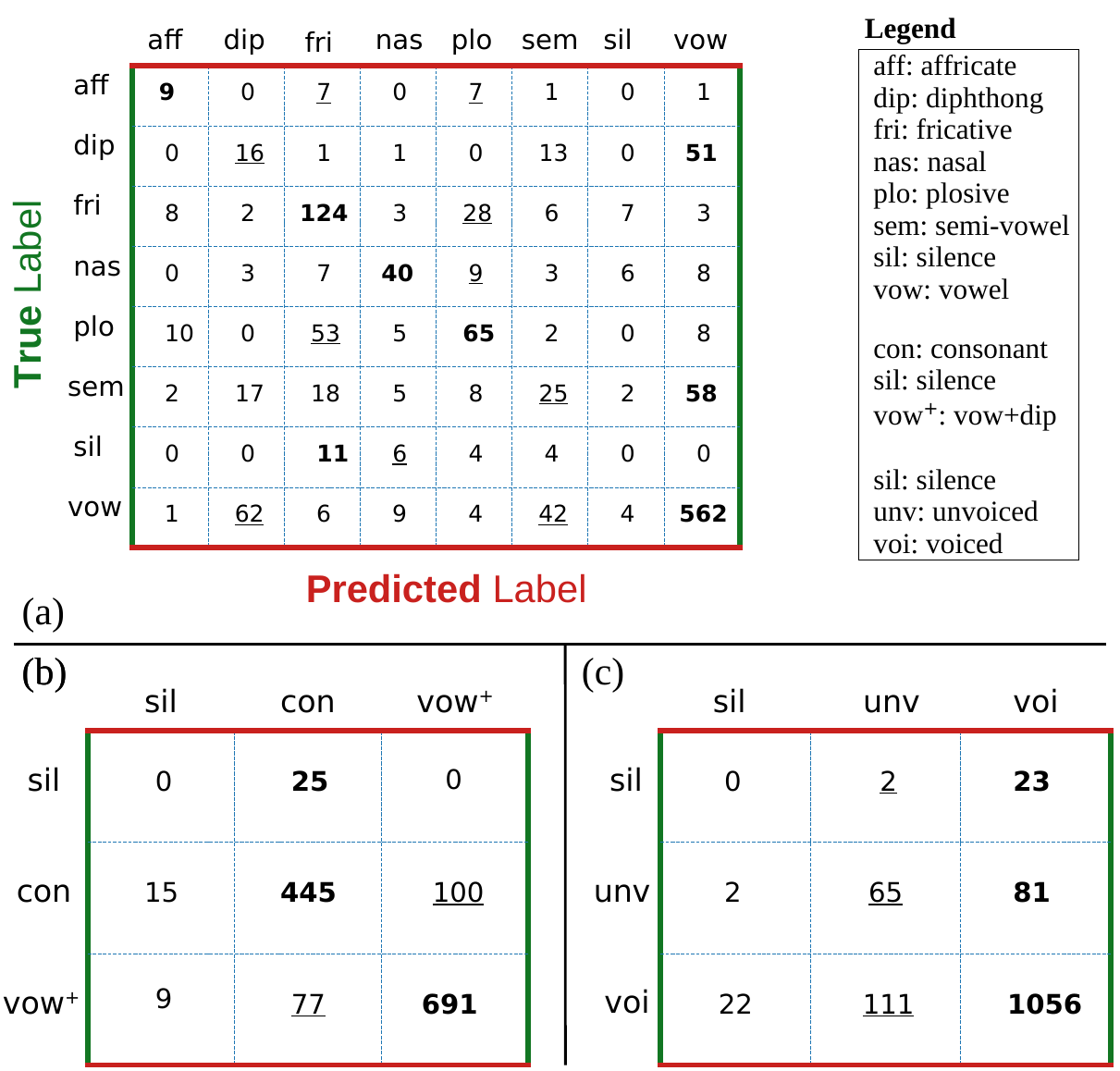}
  \caption{Confusion matrices of three phonetic categorisations for Sinc2Net on TIMIT's Dev set. The \textbf{bold} and \underline{underlined} numbers denote the first and second mostly confused classes.}
  \label{fig:conf-sinc2net}
\end{figure}

\begin{figure}[t!]
  \centering
  \includegraphics[width=\linewidth]{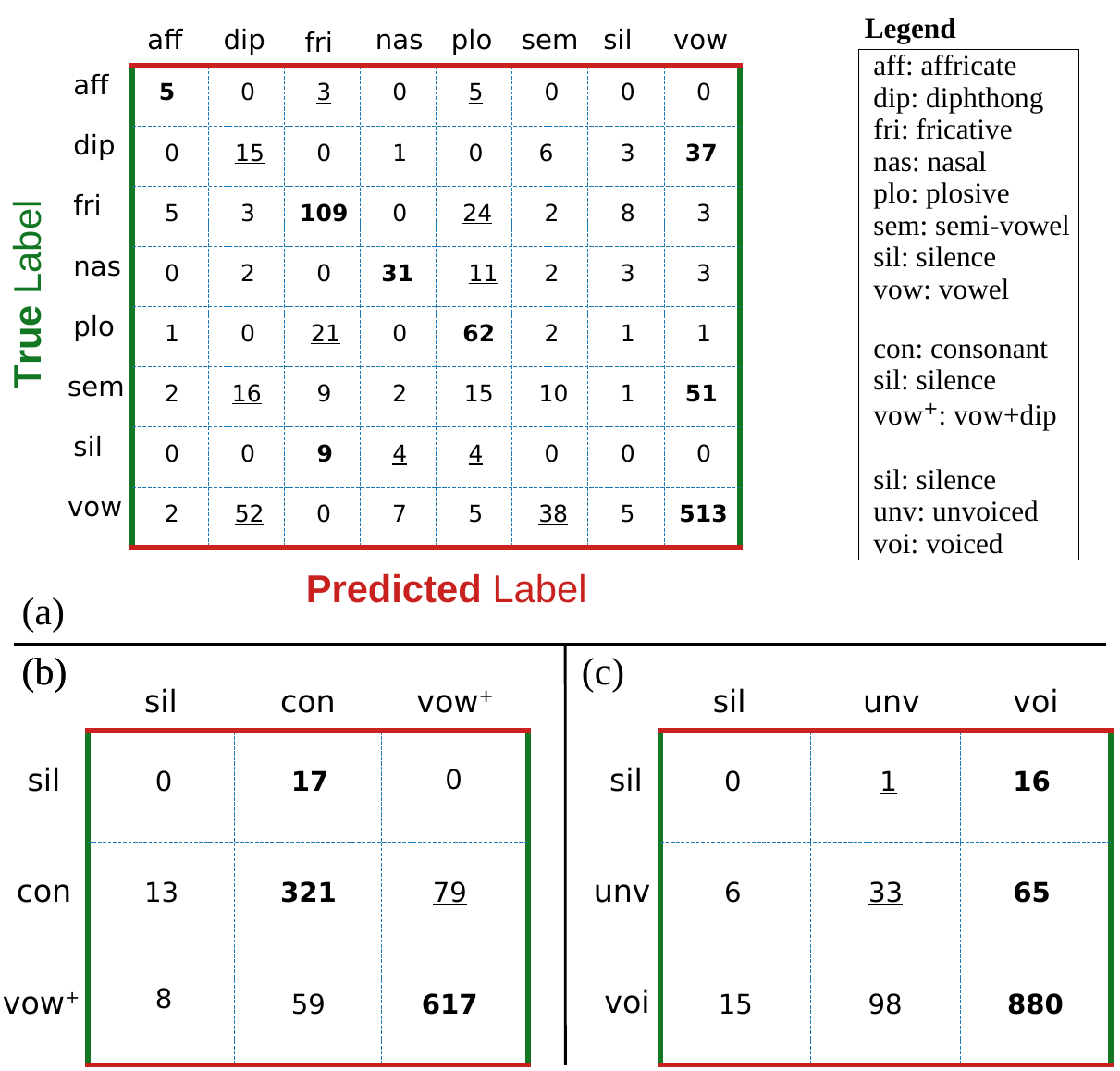}
  \caption{Confusion matrices of three phonetic categorisations for RawWav-CNN-WSJ on TIMIT's Dev set. The \textbf{bold} and \underline{underlined} denote the first and second mostly confused classes.}
  \label{fig:conf-wav-wsj}
\end{figure}

Figs.~\ref{fig:conf-sinc2net} and \ref{fig:conf-wav-wsj} present the confusion matrices on the three phonetic categorisations for the raw waveform models, without and with transfer learning from WSJ, respectively. The confusions are computed using the substitution errors: the $[i,j]$ entry of each confusion matrix reflects the number of times the phones belonging to the BPC of the $i^{th}$ row have been confused with the phones belonging to BPC of $j^{th}$ column. 

To facilitate comparison with the confusion matrices of the various systems in \cite{lea2023}, such as Wav2vec 2.0 and FBank systems, we have reported the 1${^st}$ and 2$^{nd}$ most confused classes for each system in Table~\ref{tab:conf-bpc}. Notably, there is a marked similarity in the confusion patterns among broad phonetic categories (BPCs) across different systems. For example, the Plosives and Fricatives or Vowels, Diphthongs and Semi-vowels are consistently highly confusable over all systems which is attributed to class confusability, as discussed. Another example is Affricates which are consistently confused with Fricatives and Plosives, or Nasals which are mostly confused with Plosives and Silence.

Note that the diagonal items in the confusion matrices indicate the number of within-class confusions. As seen, the diagonal element for the Silence class is always zero because it is a single-class category (Table~\ref{tab:bpc}). As such, there are no other classes in this category to cause within-class confusion.

Note that the diagonal items in each confusion matrix indicate within-class confusions. In all confusion matrices, the diagonal element for the Silence class is zero because it is a single-class category (Table~\ref{tab:bpc}), meaning there are no other classes within this category to cause within-class confusion.

Transfer learning from WSJ, has a significant effect on the performance of the raw waveform models, resulting in PERs of 11.8\% and 13.7\% on the Dev and Test sets, respectively. However, the error distribution across BPCs (Figs.~\ref{fig:psdi} and \ref{fig:cv-vu}) and confusion patterns (Table~\ref{tab:conf-bpc}) remain largely similar.

\begin{table*}[t!]
\centering
\caption{{The first and second most confused classes for various BPCs and systems on TIMIT's Dev set. For example, (vow, dip/sem) means the first most confused class is Vowel; Diphthogs and Semi-vowels are tied for the second most confused class.}}
\vspace{-2mm}
\begin{tabular}{l|c|c|c|c|c|c|c|c}
\hline
System & Affricate & Diphthong & Fricative & Nasal & Plosive & Semi-Vowel & Silence & Vowel  \\
\hline
\hline
FBank \cite{lea2023}   & aff, fri & vow, dip/sem & fri, plo & nas, plo & plo, fri & vow, sem & fri/nas, plo & vow, dip/sem \\
FBank-WSJ \cite{lea2023} & fri, aff/plo & vow, dip & fri, plo & nas, sil & plo, fri & vow, sem & nas, plo & vow, dip \\ 
Wav2vec 2.0 \cite{lea2023} & plo, sem & vow, sem/sil & fri, plo/sil & nas, sil & plo, fri/aff & vow, sem & plo, fri/sem & vow, dip \\
\hline
RawWav   & aff, fri/plo & vow, dip & fri, plo & nas, plo & plo, fri & vow, sem & fri, nas & vow, dip \\ 
RawWav-WSJ  & aff, plo & vow, dip & fri, plo & nas, plo & plo, fri & vow, dip & fri, nas/plo & vow, dip \\ 
\hline
\hline
\end{tabular}
\label{tab:conf-bpc}
\end{table*}

\begin{figure}
  \centering
  \includegraphics[width=0.87\linewidth,height=89mm]{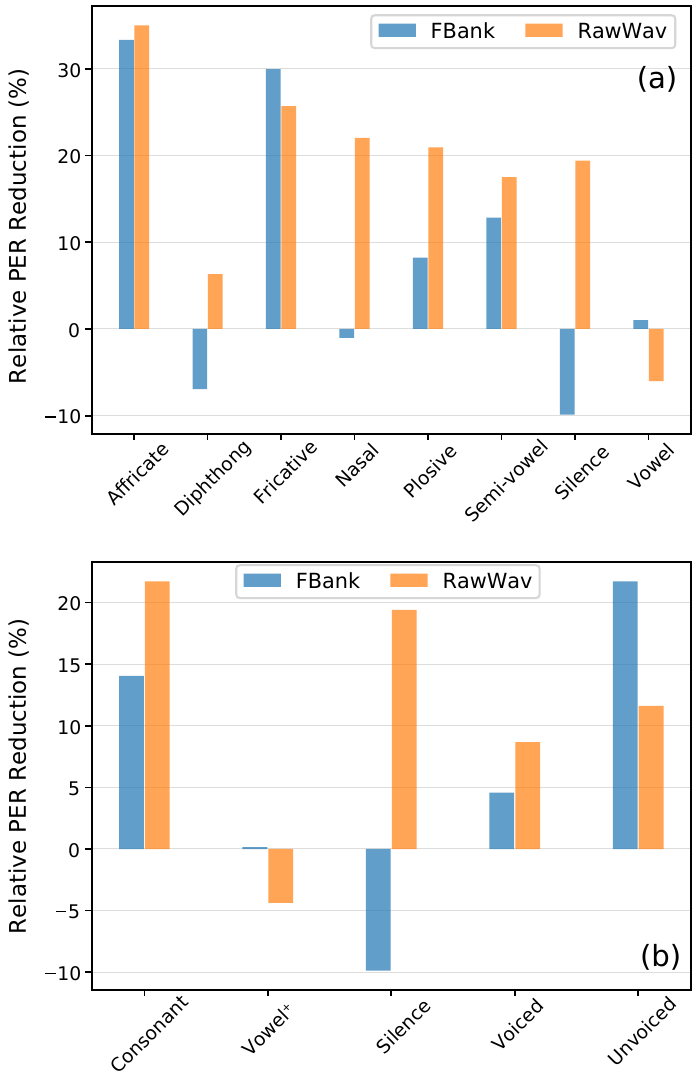}
  \caption{Relative gain after transfer learning from WSJ.}
  \label{fig:rel}
\end{figure}

Fig.~\ref{fig:rel} illustrates the performance gain (relative PER reduction) after transfer learning from WSJ across BPCs for both FBank and raw waveform models. The average PER gains shown in Fig.~\ref{fig:rel}~(a) is $8.4$\% for FBank and $17.6$\% for raw waveform models. This difference can be attributed to the raw waveform model's access to richer information, albeit requiring more data and larger model to fully leverage its potential.

There are two additional important observations in Fig.~\ref{fig:rel}. Firstly, for both FBank and raw waveform models, the performance gain after transfer learning for the Vowel$^+$ class, namely union of Vowels and Diphthongs, is minimal or even negative. These classes are particularly sensitive to speaker attributes (e.g., speaker ID \cite{Eatock1994,Margit2006} and emotion \cite{Yuan2021TheRO}). The transfer learning from WSJ does not adequately address speaker variability and speaker invariant representation learning because the WSJ training set (si284) comprises only $282$ speakers while TIMIT has a richer speaker space with $630$ speakers.

Secondly, a significant performance gap is observed for the Nasal and Silence classes between FBank and raw waveform models. While the former experiences a relative performance degradation of -1\% for Nasals and -10\% for Silence, the latter achieves an improvement of 22\% and 19\%, respectively. This observation, particularly for the Silence class, is remarkable as even advanced models like Wav2vec 2.0 struggle to enhance performance over the silence class (please refer to Fig.~24 in \cite{lea2023}).
Such a performance gap can be partially attributed to the additional information in the raw waveform, namely the phase spectrum, helping in better handling of these classes and, consequently, more effective leveraging of transfer learning.

\section{Conclusion}
In this paper, we conducted an extensive evaluation of raw waveform acoustic models on TIMIT phone recognition task, moving beyond the commonly used PER metric. 
Our analysis involved decomposing the overall substitution, deletion, insertion, and PER, calculating each metric for each broad phonetic class (BPC) within three phonetic categorisations: \{affricate, diphthong, fricative, nasal, plosive, semi-vowel, silence, vowel\}, \{consonant, vowel$^+$, silence\}, and \{voiced, unvoiced, silence\}.
We developed a raw waveform model with the highest performance on TIMIT, compared with raw waveform models reported in the literature and computed the PER for each BPC in each category. Furthermore, we examined the impact of transfer learning from WSJ on the raw waveform model's performance across various BPCs. We also constructed a confusion matrix for each phonetic categorisation, both for the raw waveform models without and with transfer learning, and compared the phonetic confusion patterns with those of the Wav2vec 2.0 and Filterbank systems.
Future research directions encompass exploring alternative modeling techniques and examining how different languages influence errors within the BPCs.


\clearpage
\bibliographystyle{IEEEtran}
{\small \bibliography{main}}

\end{document}